\documentclass[pre,floats,aps,amssymb,preprint,epsf,epsfig,rotate]{revtex4}
\usepackage{graphicx}% Include figure files
\usepackage{color}
 
\begin{document}

\title{Dynamic response of $Ag$ monolayers adsorbed on $Au(100)$ upon an 
oscillatory variation of the chemical potential: A Monte Carlo simulation study.}

%\vskip 0.5 true cm
\author{M. Cecilia Gim\'enez (1) and Ezequiel V. Albano (2).}
%\vskip 0.3 true cm

\affiliation{(1) Laboratorio de Ciencias de Superficies y Medios Porosos,
Departamento de F\'{\i}sica,
Facultad de Ciencias F\'{\i}sico-Matem\'aticas y Naturales,
Universidad Nacional de San Luis, CONICET,
Chacabuco 917, 5700, San Luis, Argentina.\\
(2)Instituto de Investigaciones Fisicoqu\'{\i}micas Te\'{o}ricas
y Aplicadas, (INIFTA), CONICET, UNLP.
Sucursal 4, Casilla de Correo 16, (1900) La Plata. ARGENTINA. 
FAX : 0054-221-4254642.
E-mail : ealbano@inifta.unlp.edu.ar}

\date{\today}

%\newpage

\begin{abstract}
Based on the fact that the underpotential electrochemical deposition 
of Ag atoms on the $Au(100)$ surface exhibits sharp first-order 
phase transitions at well-defined values of  the (coexistence) 
chemical potential ($\mu_{coex}$), we performed 
extensive simulations aimed at investigating the hysteretic 
dynamic behavior of the 
system close to coexistence upon the application of a 
periodic signal of the form 
$\mu(t) = \mu_{coex} + \mu_{o}sin(2\Pi t/\tau) $, where  
$\mu_{o}$ and $\tau$ are the 
amplitude and the period of the sweep, respectively. 
For relatively short periods and small 
enough amplitudes the system becomes trapped either 
at low or high Ag coverage states, 
as evidenced by recording hysteresis loops. This 
scenario is identified as dynamically 
ordered states (DOS), such that the relaxation time $(\tau_{relax})$ 
of the corresponding metastable state obeys $\tau_{relax} > \tau $. 
On the other hand, by properly increasing 
$\mu_{o}$ or/and $\tau$, one finds that the $Ag$ coverage 
gently follows the external drive (here $\tau_{relax} < \tau $) 
and the system is said to enter into  
dynamically disordered states (DDS), where hysteresis loops 
show almost symmetric limiting cycles. This symmetry breaking 
between limiting cycles driven by an external 
signal is discussed in terms of the concept of (out-of-equilibrium) 
Dynamic Phase Transitions between DOS and DDS, similar to 
those encountered when a magnetic system is 
placed in the presence of a variable external magnetic field.
However, a careful finite-size scaling study reveals that, 
at least at $T = 300K$,  the 
$Ag/Au(100)$ system does not exhibit true second-order
phase transitions but rather a crossover 
behavior between states.
A diagram showing the location of the ordered and disordered states
in the $\mu$ versus $\tau$ plane is obtained and discussed. 
\vskip 0.5 true cm



\end{abstract}
\maketitle

\pagebreak

\section{INTRODUCTION}

The term hysteresis is used to describe the lagging of an effect
behind its cause. Hysteresis is a common phenomenon that is
observed in a great variety of physical,
chemical, and biological systems. However, the magnetization
response of a ferromagnet in the presence of an oscillatory
magnetic field is probably the best known example of
hysteresis.

Hysteresis in thermodynamic systems is often related to the
existence of first-order phase transitions, which are the sources
of nonlinearities always associated with hysteretic behavior.
Within this context, systems exhibiting first-order phase
transitions and capable of becoming coupled to an external
oscillatory drive are excellent candidates for the observation of
(out-of-equilibrium) dynamic phase transitions (DPT's). A DPT takes
place between a dynamically ordered state (DOS) and a dynamically
disordered state (DDS) when the drive changes from high to low
frequencies. Most studies related to DPT's have been performed by
using magnetic systems. Particularly, the Ising ferromagnet on a
two-dimensional square lattice has become the subject of great
interest \cite{dro1,dro2,dro3,chakra1,chakra2}. In this case, an oscillatory 
magnetic field causes the
occurrence of a second-order DPT that is believed to
belong to the universality class of the standard Ising model in
the same dimensionality \cite{dro1,dro2,dro3}. For a recent review see
reference \cite{acha}.

Aside from ferromagnets, sharp first-order transitions
are also observed in other physical systems of relevant practical
importance, namely upon the underpotential electrochemical
deposition (Upd) of metal atoms on metallic surfaces. Upd
implies the deposition of a metal on the surface of an electrode
of different nature at potentials more positive than those
expected from Nernst 
equation \cite{Kolb,a,b,Leiva_UPD_2,Lorenz_2004}. Therefore, this
kind of phenomena can be intuitively understood by considering 
that, under Upd conditions, it is more favorable for the adsorbate 
to be deposited on a substrate of different nature than 
on a surface of the adsorbate itself.
For a recent review on this subject
see e.g. the book by Budevski et al \cite{a} while for the
discussion of theoretical aspects see e.g. \cite{b}.

The interpretation of Upd as a phase transition has been employed
by Blum et al \cite{c} in order to model the underpotential
deposition of Cu on Au. Subsequently, the concept has been
generalized by Staikov et al \cite{d}. The formation and growth of
well ordered overlayers can be understood as a first-order phase
transition that takes place at the electrode-solution interface.
This kind of transition can be detected by observing a
discontinuity in the adsorption isotherm that is
accompanied by a very sharp peak of the corresponding
voltammetric profile \cite{c}. It should also be mentioned that
this oversimplified picture is not completely free of controversy
and is the subject of current debate \cite{a}. Computer
simulations performed by one of us \cite{e} have recently contributed to
the understanding of the deposition of Ag on Au (100), where sharp
first-order transitions in the adsorption isotherms have been
reported.

Within this context, the aim of the present paper is to study the
dynamic behavior of the Ag/Au(100) electrochemical system by means
of computer simulations. Taking advantage of our experience in the
simulation of the equilibrium properties of such system, here we
propose investigating its response to periodic variations of the
chemical potential. This situation can experimentally be
achieved just by changing the applied potential. Of course, the
effect of oscillatory variations of different frequency will be
studied and discussed.

The organization of this paper is as follows: in Section II we
describe the model and the numerical simulation technique. Also,
some relevant results corresponding to the system Ag/Au(100) under
equilibrium are discussed. The obtained  results are
presented and discussed in Section III. Our conclusions are
stated in Section IV where some experiments that could account for
the numerical simulations are suggested.

\section{Description of the model and the simulation method.}

Monte Carlo simulations are performed assuming a lattice model,
with periodic boundary conditions, to represent the square Au(100)
substrate. Each lattice site represents an adsorption site for a Ag
atom. It may absorb or desorb. 
If the crystallographic misfit between the involved atoms is not important,
it is a good approximation to assume that the adatoms adsorb on defined
discrete sites on the surface, given by the positions of the substrate
atoms. This is the case of the very well studied  $Ag/Au(100)$ system
\cite{Ag/Au_Garcia, Ag/Au_Marijo, Ag/Au_Garcia2, Ag/Au_Hara}.
The Grand Canonical ensemble is used throughout and the 
transitions rates are evaluated according to  
the Metropolis algorithm. That is, for a given temperature 
($T$) and chemical potential ($\mu$), a  site is selected at random an 
its state is changed, becoming either empty or occupied 
by $Ag$ atoms, with probability

\begin{equation}
W=min[1,exp(-(\Delta E - \mu \Delta N)/k_B T)] ,
\end{equation}

\noindent where $k_B$ is the Boltzmann constant, $\Delta N$ is $+1$ or
$-1$ depending of the change of occupation (adsorption or desorption)
and $\Delta E$ is equal to $E_{ads}$ or $-E_{ads}$. These
adsorption energies have previously been  calculated by 
employing the embedded atom model (see below) considering the  
environment of the involved site, taking into account first, second 
and third neighbors. A careful discussion of the
assumptions involved in the lattice model in the light of the
available experimental information has already been performed
\cite{e} and does not need to be repeated here.

A very important feature of the used method is that the
interatomic potentials involved in the Monte Carlo simulations are
first evaluated by means of the embedded atom model (EAM). The EAM
assumes that the total energy of an arrangement of N atoms may be
calculated as the sum of individual energies of the atoms. Thus, the
attractive contribution to the EAM potential is given by the
embedding energy, which accounts for many-body effects. On the
other hand, the repulsion between ion cores is accounted for pair
potentials depending on the distance between cores only. For
additional details see reference \cite{e} and references therein.
For the purpose of the present work, the EAM has been
parametrized to fit available experimental data such as elastic
constants, dissolution enthalpies of binary alloys, bulk lattice
constants and sublimation heats \cite{f}.

The Grand Canonical Monte Carlo method allows us to take the
chemical potential ($\mu$) as an independent variable that
can straigthforwardly be related to the electrode potential that
is used to control the chemical potential of the species at the
metal/solution interface. 
Also, in our two-dimensional
system of size  $L \times L = M$, a Monte Carlo step (mcs)
involves the random selection of $M$ adsorption sites in order to
consider all possible adsorption/desorption events. In these
simulations, diffusion events are not considered explicitly, but
instead one has an equivalent process since there exists the 
possibility that an atom may desorb from one site
while another one may adsorb on a neighboring site.
It should be mentioned that very recently Frank et al \cite{Fra} 
have studied the influence of lateral adsorbate diffusion
on the dynamics of the first-order phase transition within the 
framework of the two-dimensional Ising model. These results are 
discussed in the context of Upd and one relevant finding 
is that diffusion may suppress the nucleation processes around
clusters.

Simulations are performed starting from an empty substrate and
subsequently the chemical potential is swept harmonically
according to

\begin{equation}
\mu(t) = \mu_{coex} + \mu_{o} \sin(\omega t) \label{mudete}
\end{equation}

\noindent where  $\mu_{o}$ is the amplitude of the sweep, while
$\omega = 2\Pi/\tau$ is the pulsation such that $\tau$ is the period
of the applied perturbation. Also, $\mu_{coex}$ is the  coexistence
chemical potential meassured at the first-order phase transition
observed upon deposition of $Ag$ on $Au(100)$ (see e.g. figure
\ref{Fig1} and the discussion below).
During the simulation we record the time dependence of the
Ag-coverage ($\theta(t)$). After neglecting several initial sweeps
in order to allow the system to achieve a nonequilibrium
stationary state, we calculate the quantities of interest by
averaging over many cycles.

\section{Results and Discussion.}

\begin{figure}
\centerline{
\includegraphics[width=10cm,height=10cm,angle=0]{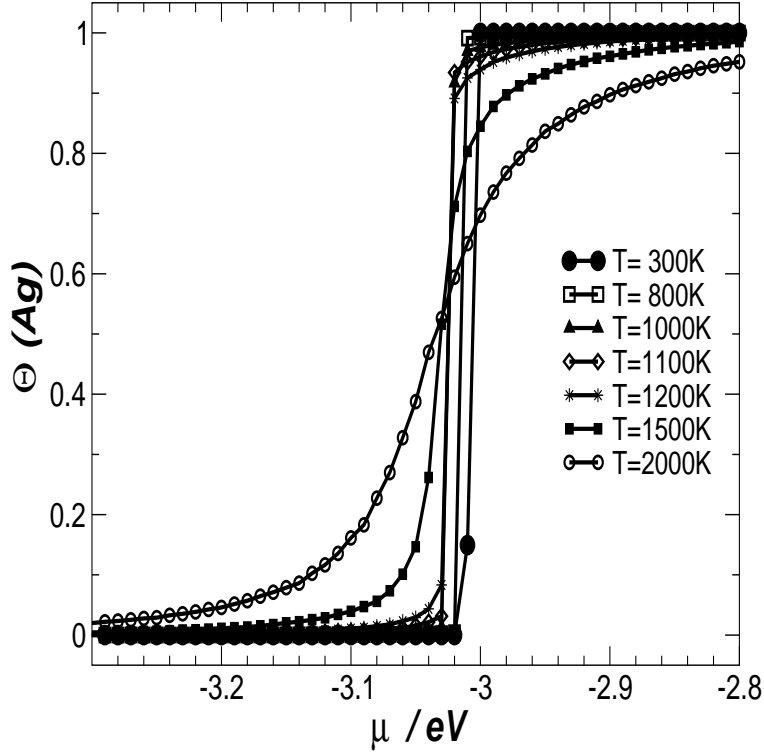}}
\caption{Adsorption isotherms for the deposition of Ag
on Au(100) obtained under equilibrium conditions by means of
numerical simulations. Results corresponding to different
temperatures as listed in the figure. For further details see
reference \cite{e}.}
\label{Fig1}
\end{figure}

Before starting the discussion of the dynamic behavior of the
system, it is instructive to remind the reader that both experiments and 
numerical simulations 
performed under equilibrium conditions have confirmed that the
adsorption isotherms ($\theta_{Ag}$ {\it vs} $\mu$) exhibit
first-order transitions within a wide range of temperatures, as
is shown in figure \ref{Fig1} for the case of Monte Carlo results 
performed by one of us \cite{e}.

In view of these results we performed the study of the
dynamic behavior of the system by taking T=300K, so
that the system is far below its critical temperature. At this
temperature one has that the chemical potential at coexistence is
$\mu_{coex} = -3.03 eV$. Therefore, the sweep of the chemical
potential, as given by equation (\ref{mudete}), is applied over
$\mu_{coex}$.
%, so that the actual value of the chemical potential
%is given by $\mu^{a}(t) = \mu_{coex} + \mu(t)$.

\begin{figure}
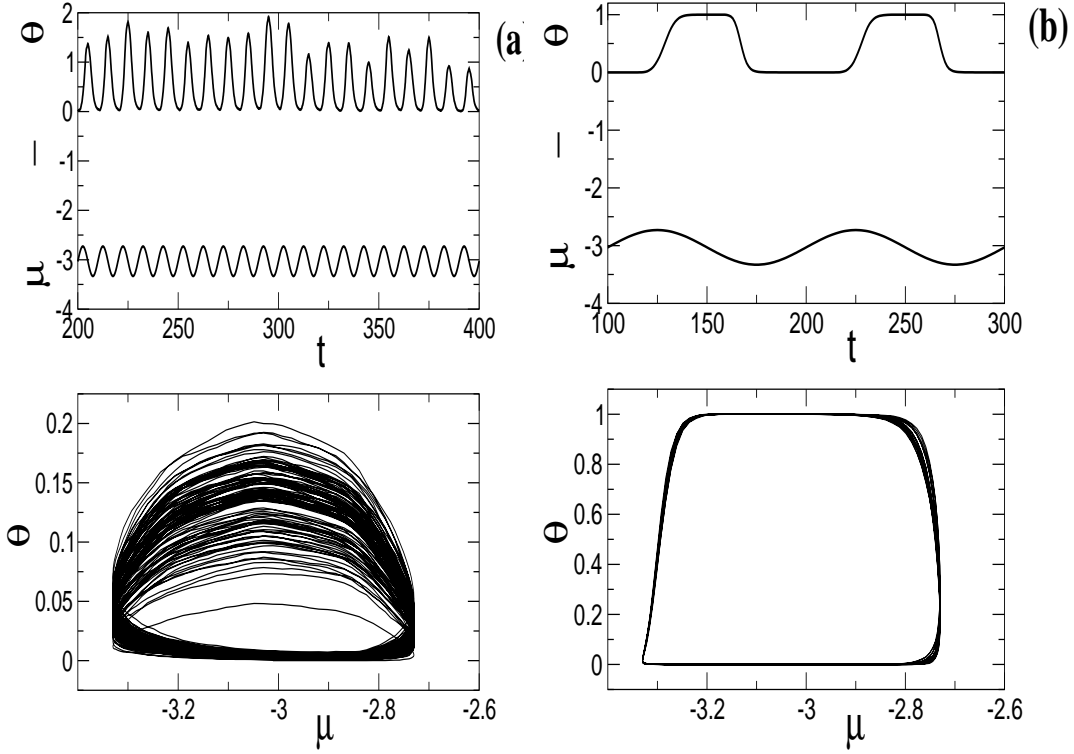

\centerline{
\includegraphics[width=7cm,height=10cm,angle=0]{Fig2a.eps}
\includegraphics[width=7cm,height=10cm,angle=0]{Fig2b.eps}}
\caption{Plots of the temporal dependence of the coverage and the
chemical \textbf{potential} (upper panel), and the
coverage-chemical potential loops (lower panel). (a) Results
obtained using lattices of side $L = 100$ and by taking $\mu_{o} =
0.3$ and $\tau = 10$. (b) As in (a) but for $\tau = 100$. In (a)
the coverage is amplified by a factor $10$ for the sake of clarity.
So, in order to properly obtain the actual coverage the 
$\theta-$scale has to
be divided by a factor $10$.}
\label{Fig2}
\end{figure}

Figure \ref{Fig2} shows two examples of the dynamic behavior of the
system as obtained by applying oscillatory sweeps of the same
amplitude $\mu_{0} = 0.3$ but different periods $\tau$.

For the shortest period ($\tau = 10 mcs$, see figure
\ref{Fig2}(a)) the system becomes trapped within a low-coverage
regime, with $\theta_{Ag} < 0.5$. Since the highest achieved
chemical potential is given by $\mu_{coex} - \mu_{0} =
-2.73 eV$, the applied signal clearly drives the system well inside
the high-coverage regime of the equilibrium isotherm (see figure
\ref{Fig1}). So, the behavior observed in figure \ref{Fig2} for
$\tau = 10$ is clear evidence that such a period is much smaller than
the relaxation time ($\tau_{relax}$) required by the system to
jump from the low-coverage to the high-coverage regimes. On the
other hand, for $\tau = 100$ (see figure \ref{Fig2}(b)) the system
reaches the high-coverage regime ($\theta_{Ag} \simeq 1$) during
all sweeps, indicating that for this case one has $\tau >
\tau_{relax}$. So, from figure \ref{Fig2} one concludes that it is
possible to identify the dynamic competition between two time
scales in the system: the half-period of the external drive and
the relaxation time (also known as the metastable lifetime of the
system in a given state). For large periods, a complete decay of
the metastable phase always occurs during each half-period.
Consequently, the coverage describes an (almost symmetric) limiting
cycle. In contrast, for short periods, the system does not have
enough time to change the coverage from $\theta \cong 0$ to
$\theta \simeq 1$ and the symmetry of the hysteresis loop is
broken.

It should be mentioned that the
phenomenon of symmetry breaking between limiting cycles
in an externally driven system has been the subject of considerable
attention. It was first reported in the context of numerical and
mean-field studies of the magnetization of a ferromagnet in an
oscillating magnetic field \cite{a1,a2} and subsequently it
has been studied by means of Monte Carlo simulations of the
kinetic Ising model \cite{dro1,dro2,dro3,chakra1,acha}.

In order to study the DPT's involved in the already discussed symmetry
breaking process in magnetic systems it is useful to define the
dynamic order parameter ($Q$) as the period-averaged
magnetization. Since in the present work we are interested in the
surface coverage with Ag atoms, we take advantage of the lattice
gas-spin system equivalence. In fact, if the spin variable can
assume two different states ($s = \pm 1$) corresponding to two
occupation states of lattice gas ($n = 0$ (empty)) and $n = 1$ 
(occupied)), one has that  $s = 2n-1$ and the suitable order
parameters is

\begin{equation}
Q = \frac{1}{\tau} \oint (2\theta_{Ag} - 1)dt .
\label{orpa}
\end{equation}

Notice that for finite systems, as in the present study, one
actually computes $\langle |Q| \rangle$, where $\langle
\rangle$ means averages over different cycles of the time series
$\theta_{Ag}(t)$. So, if $\tau < \tau_{relax}$ the coverage 
cannot change from $\theta \sim 0$ to $\theta \sim 1$ (and {\it
vice versa}) within a single period, and therefore one has that
$\langle |Q| \rangle > 0$. This situation is regarded as the DOS. 
In contrast, when $\tau > \tau_{relax}$ and the
coverage follows the applied chemical potential one has that $Q
\approx 0$ in the so called DDS. Between
these two extreme regimes, one expects that the existence of a critical
period such as $\langle |Q| \rangle$ should vanish in the
thermodynamic limit. This behavior may be the signature of a
nonequilibrium DPT.

Further insigth into the nature of DPT's can be obtained by measuring the 
fluctuations of the order parameter ($\chi$), given by

\begin{equation}
\chi = L^{2} Var (|Q|) = L^{2} [ \langle |Q|^{2} \rangle -  
\langle |Q| \rangle ^{2} ] ,
\label{chiorpa}
\end{equation}

\noindent where $Var (|Q|) $ is the variance of the order parameter.  
This measurement is motivated by the fact that, as is well known,  
systems undergoing second-order phase transitions exhibit a divergency 
of the susceptibility. 
Of course, for equilibrium systems the fluctuations of the order parameter are 
related to the susceptibility through the fluctuation-dissipation 
theorem (FDT). It is not obvious if such a theorem would hold for our 
out-of-equilibrium dynamic system. 
However, if our system obeyed the FDT, both quantities would be proportional. 

\begin{figure}
\centerline{
\includegraphics[width=10cm,height=10cm,angle=0]{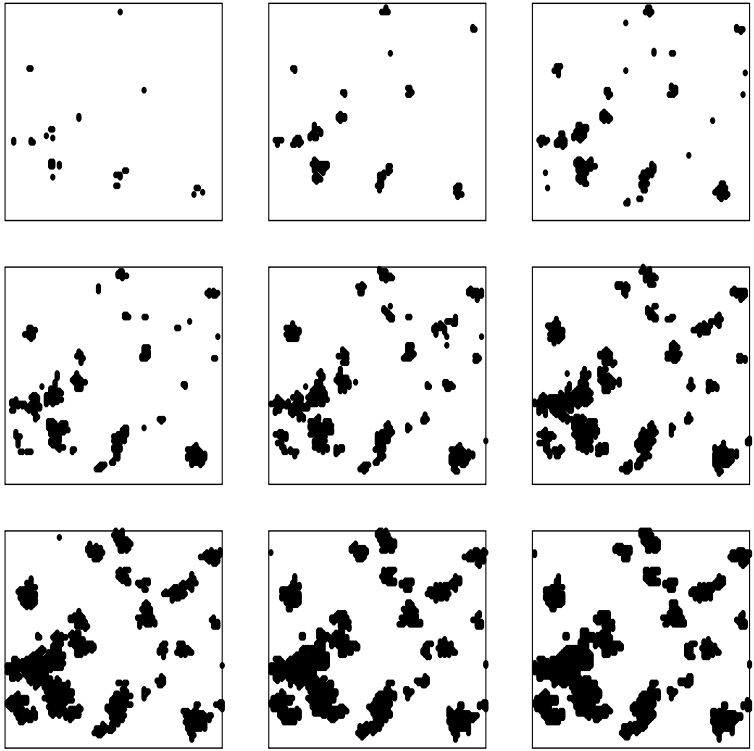}}
\caption{Typical snapshot configurations obtained during the
\textbf{adsorption} half-\textbf{cycle}. For the sake of
comparison, relevant parameters are taken as in figure
\ref{Fig2}(a), namely $\mu_{o} = 0.3$ and $\tau = 10$. The
chemical potential increases from top-left ($\mu = \mu_{coex}= -3.03$)) 
to bottom-right ($\mu = -2.73$)  .}
\label{Fig3}
\end{figure}

\begin{figure}
\centerline{
\includegraphics[width=10cm,height=10cm,angle=0]{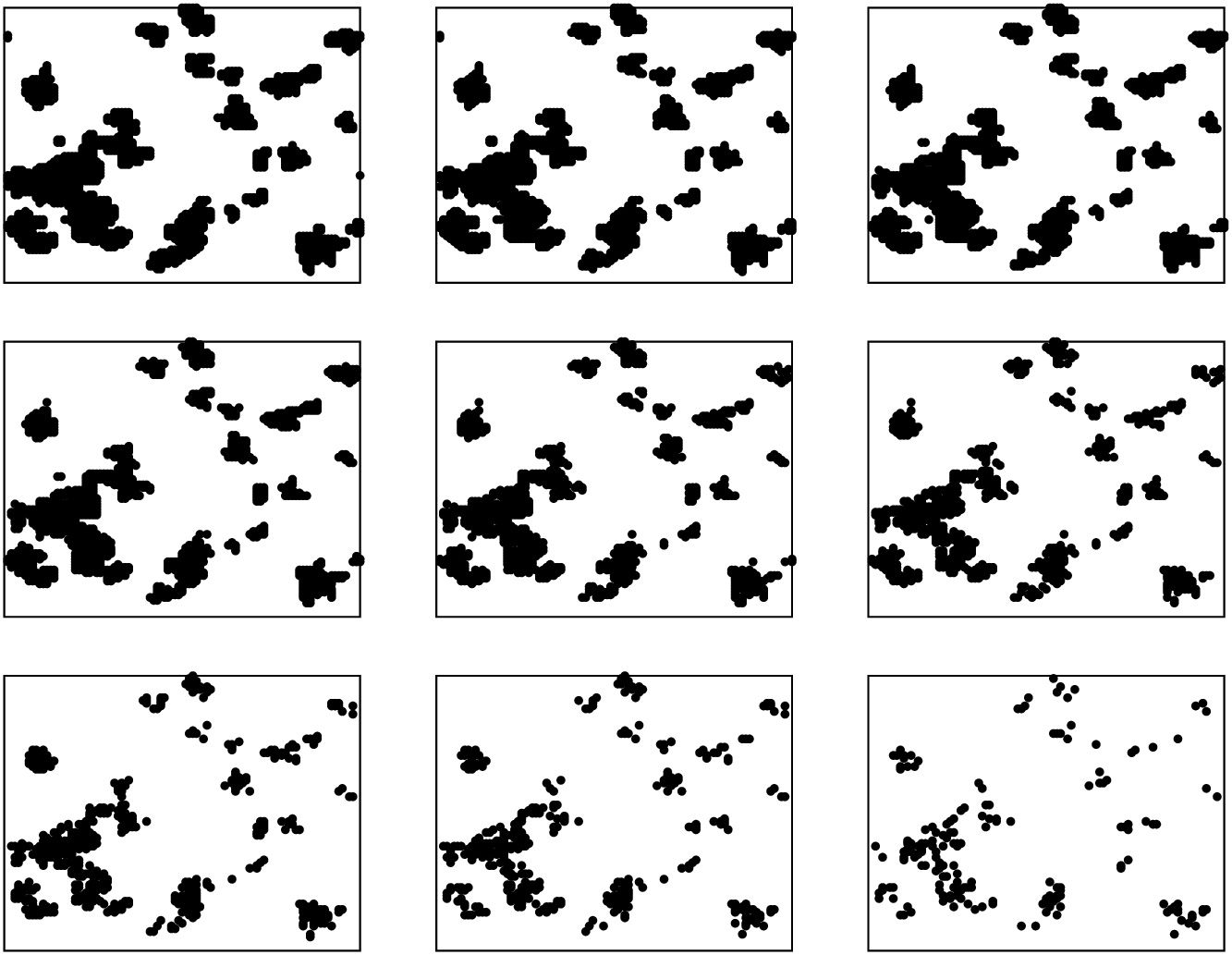}}
\caption{Typical snapshot configurations obtained during the
desorption half-\textbf{cycle}. For the sake of comparison,
relevant parameters are taken as in figure \ref{Fig2}(a), namely
$\mu_{o} = 0.3$ and $\tau = 10$. The chemical potential decreases
from top-left ($\mu = \mu_{coex}= -3.03$)) to 
bottom-right ($\mu = -3.33$)).} \label{Fig4}
\end{figure}

Figures \ref{Fig3} and \ref{Fig4} show typical snapshot
configurations obtained during the adsorption and desorption
half-cycles of the sweep, respectively. Recall that all
parameters are set as in figure \ref{Fig2} (a)  for the
sake of comparison. Figure \ref{Fig3} suggests that by 
sweeping the chemical potential, the decay
of the metastable (with $\theta_{Ag} = 0$) phase takes place by
random homogeneous nucleation of many critical clusters of Ag
(recall that for the applied chemical potential the stable phase
corresponds to a fully covered Ag surface). These critical
droplets grow and coalesce driving the system into the stable
phase, resulting in an almost deterministic process. This kind of
behavior is very well documented for the case of the Ising model
and the outlined mechanism is known as a multiple droplet (MD)
regime \cite{dro1,dro2,dro3}. In fact, in the Ising magnet the
decay of the metastable phase under a sudden reversal of the
applied magnetic field may proceed according to different
mechanisms, depending on the magnitude of the
field, the temperature $T$, and the lattice size $L$. The system
crosses over from the MD to a single droplet (SD) regime
at a crossover field called the dynamic spinodal. Within the SD
regime the decay of the metastable phase occurs by random
homogeneous nucleation of a single critical droplet of the stable
phase. For a detailed discussion of the different decay modes of
the Ising system see references \cite{dro1,dro2,dro3}. Roughly
speaking, the MD regime is observed when both the system size and
the amplitude of the applied oscillatory field are large enough
\cite{dro3}. We found that our simulations of the Ag/Au(100)
system correspond to the MD regime, presumably due to the
application of large amplitude signals ($\mu_{o} \geq 0.1$), 
the use of relatively large lattices ($L \geq 100$), and the 
relatively low temperature ($T = 300 K$) used in the simulations. 
However, we expect that the system may also display a SD regime for
appropriate selections of the parameters, but undertaking
systematic studies on this issue is beyond the scope of the present
paper.

On the other hand, figure \ref{Fig4} shows that the desorption
process involves the detachment of individual particles
from all clusters formed upon adsorption. So, in this
sense desorption and adsorption are also MD processes.
Also, the very low rate of nucleation of $Ag-$free regions
within $Ag-$clusters, observed in figure \ref{Fig4}, is in 
agreement with the fact that due to the interactions, 
adsorption (figure \ref{Fig3}) and desorption (figure \ref{Fig4}) 
are not be symmetric processes (the model lacks of symmetry under the 
exchange empty$\longleftrightarrow$ocupied sites). 

In view of these results suggesting the existence of DPT's in the 
$Ag/Au(100)$ system we  performed extensive numerical 
simulations in order to study the dependence of the order 
parameter (see equation (\ref{orpa})) and its fluctuations 
(see equation (\ref{chiorpa})) on both the chemical potential 
and the period of the applied signal, as shown in figures 5 and 6, 
respectively. For a given period and low enough 
values of $\mu_{o}$ one has that $\langle |Q| \rangle  > 0 $, 
although it decreases smoothly and monotonically when $\mu_{o}$ is 
increased (see figure 5(a)). On the other hand, by fixing 
the chemical potential, the order parameter also decreases 
when $\tau$ is increased (see figure 6(a)). 
It is worth mentioning that figure 6(a) is qualitatively simmilar
to figure 15 of the paper of Kornis et al \cite{kor},
that corresponds to the MD process and has been obtained
during the study of DPT's in the Ising model in two dimensions. 
So, the behavior 
of $\langle |Q| \rangle $ suggests the existence 
of continuous transitions between a DOS (for $\langle |Q| \rangle  > 0 $ ) 
and a DDS (for $\langle |Q| \rangle  \simeq 0 $ ). 
This preliminary conclusion, which has already been anticipated 
within the context of discussion of figure \ref{Fig2}, 
may also be supported by the characteristic peaks observed
by plotting the variance of the order parameter, 
as shown in figures 5(b), 6(b), and 6(c) .  

\begin{figure}
\centerline{
\includegraphics[width=7cm,height=10cm,angle=0]{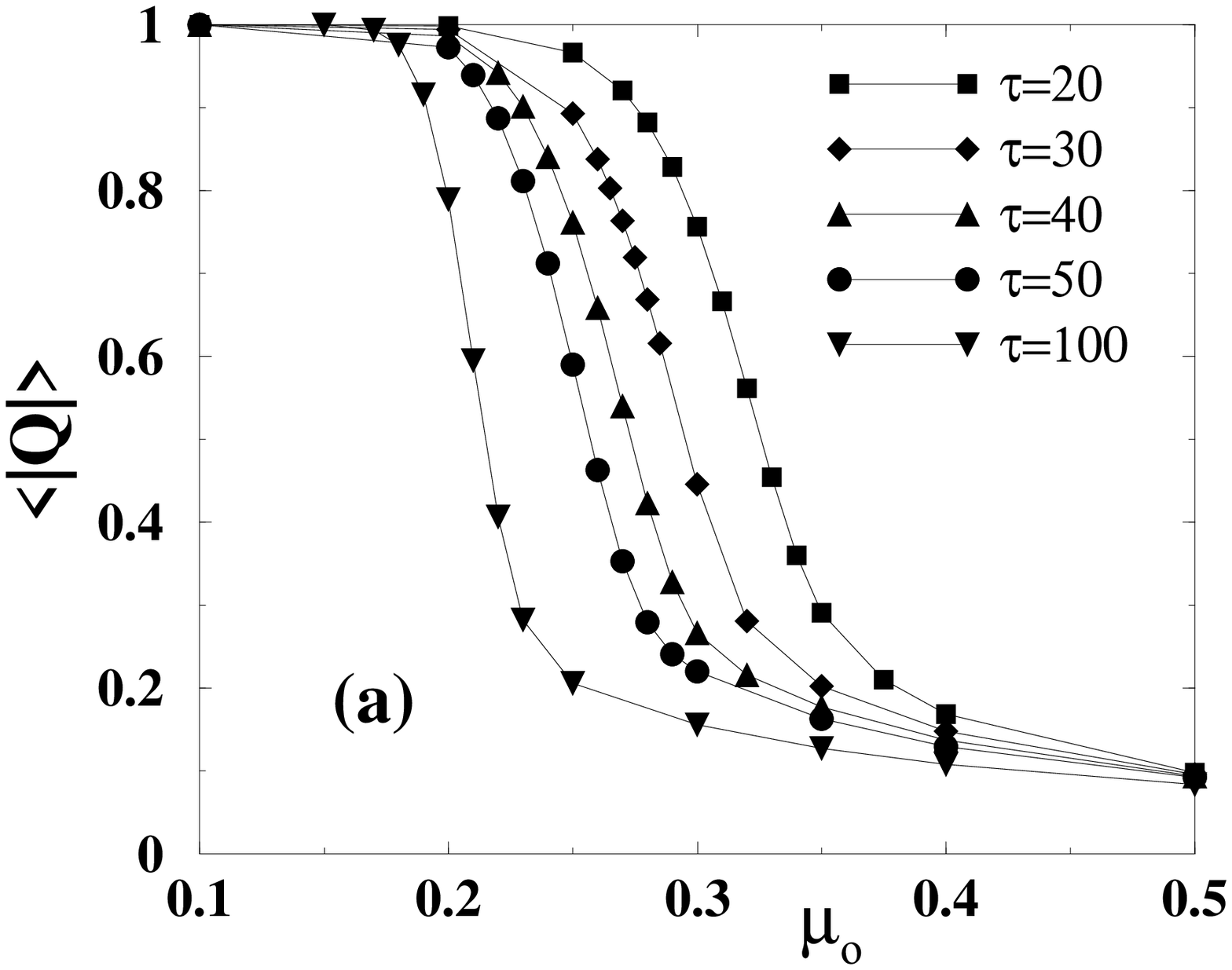}
\includegraphics[width=7cm,height=10cm,angle=0]{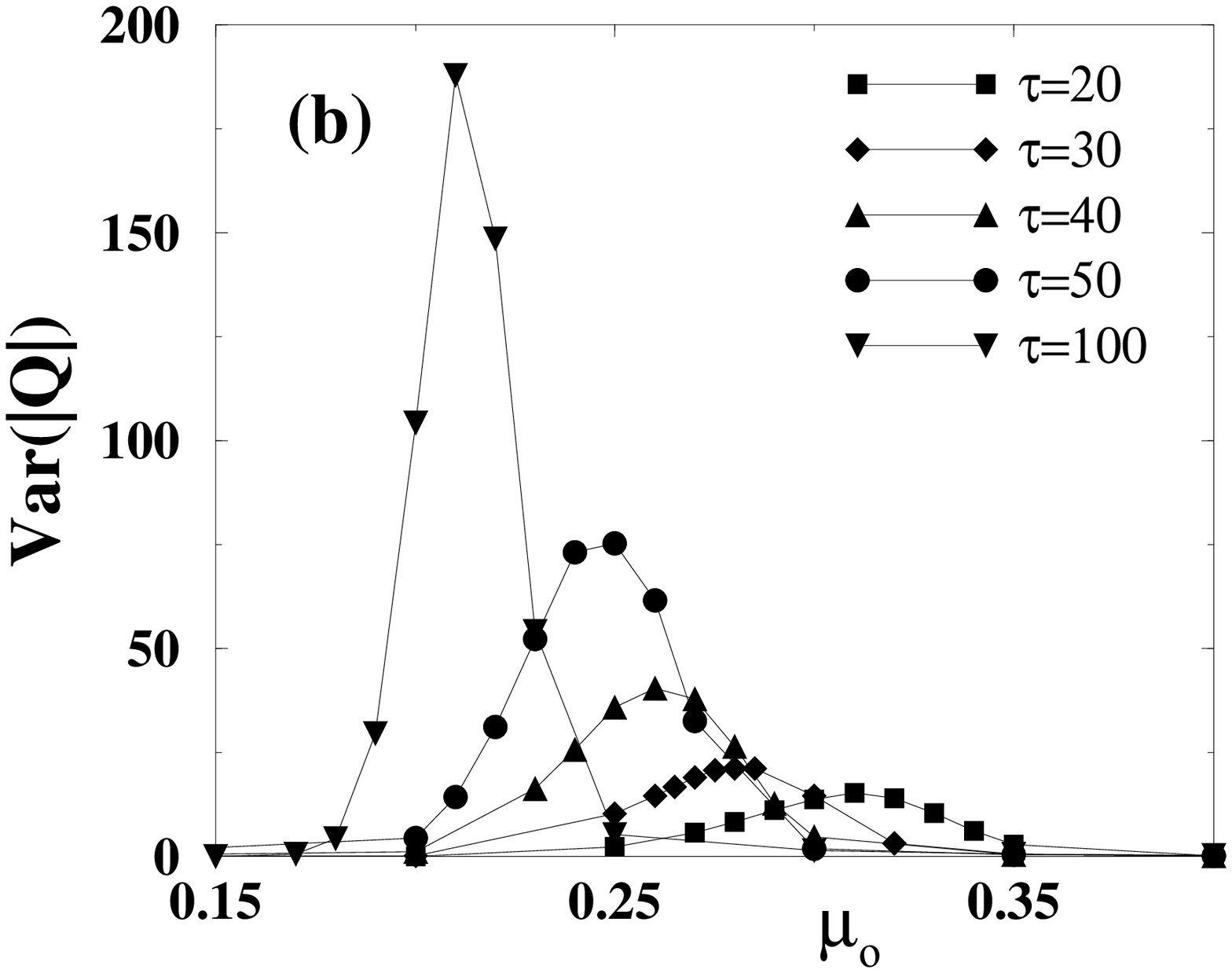}}
\caption{Plots of the (a) order parameter and (b) the fluctuations
of the order parameter versus the amplitude of the applied
(oscillatory) chemical potential, respectively. Results obtained
by using lattices of side $L = 100$ and by taking different
\textbf{periods}  $\tau$ of the applied signal, as listed in the
figures.} \label{Fig5}
\end{figure}

\begin{figure}
\centerline{
\includegraphics[width=7cm,height=10cm,angle=0]{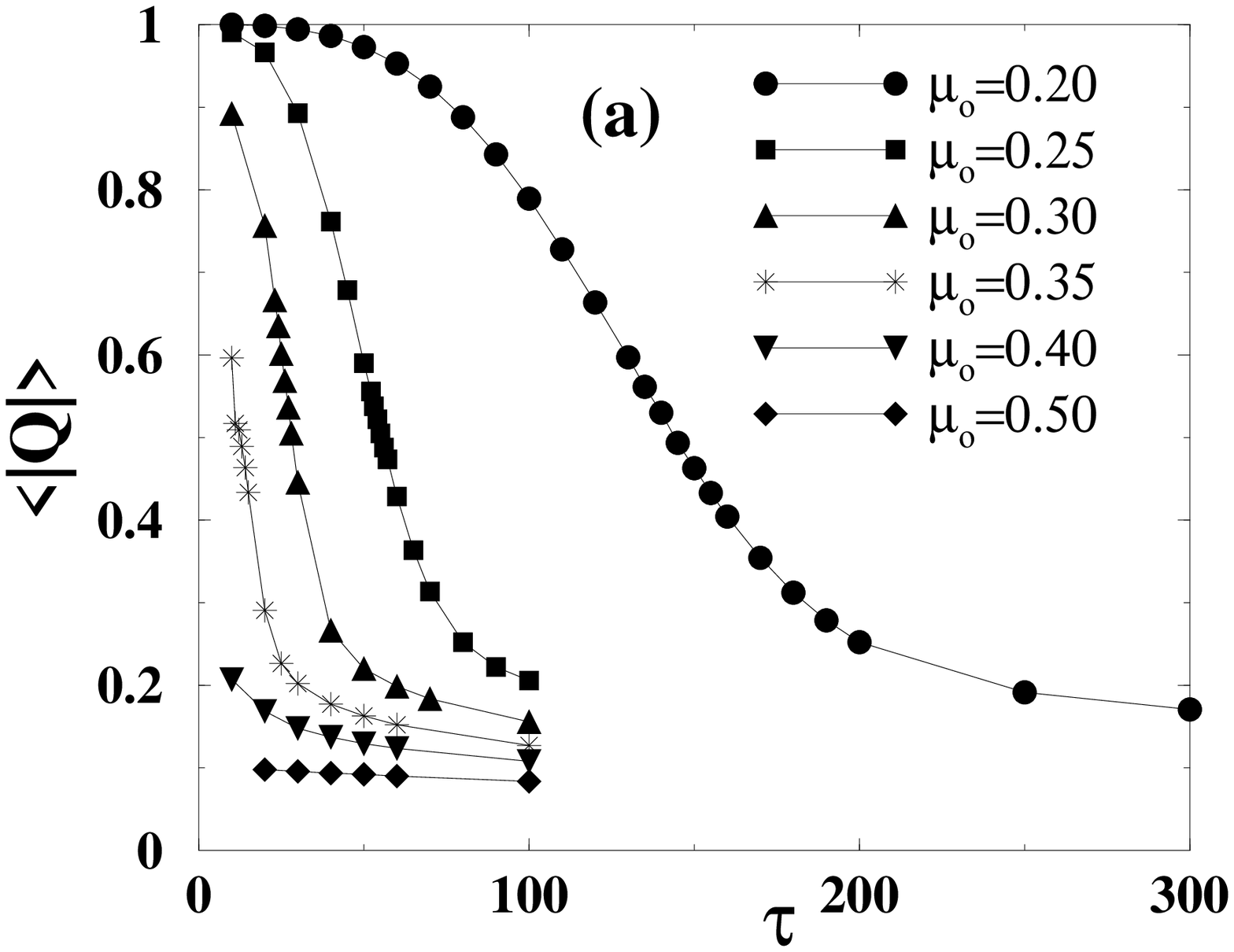}
\includegraphics[width=7cm,height=10cm,angle=0]{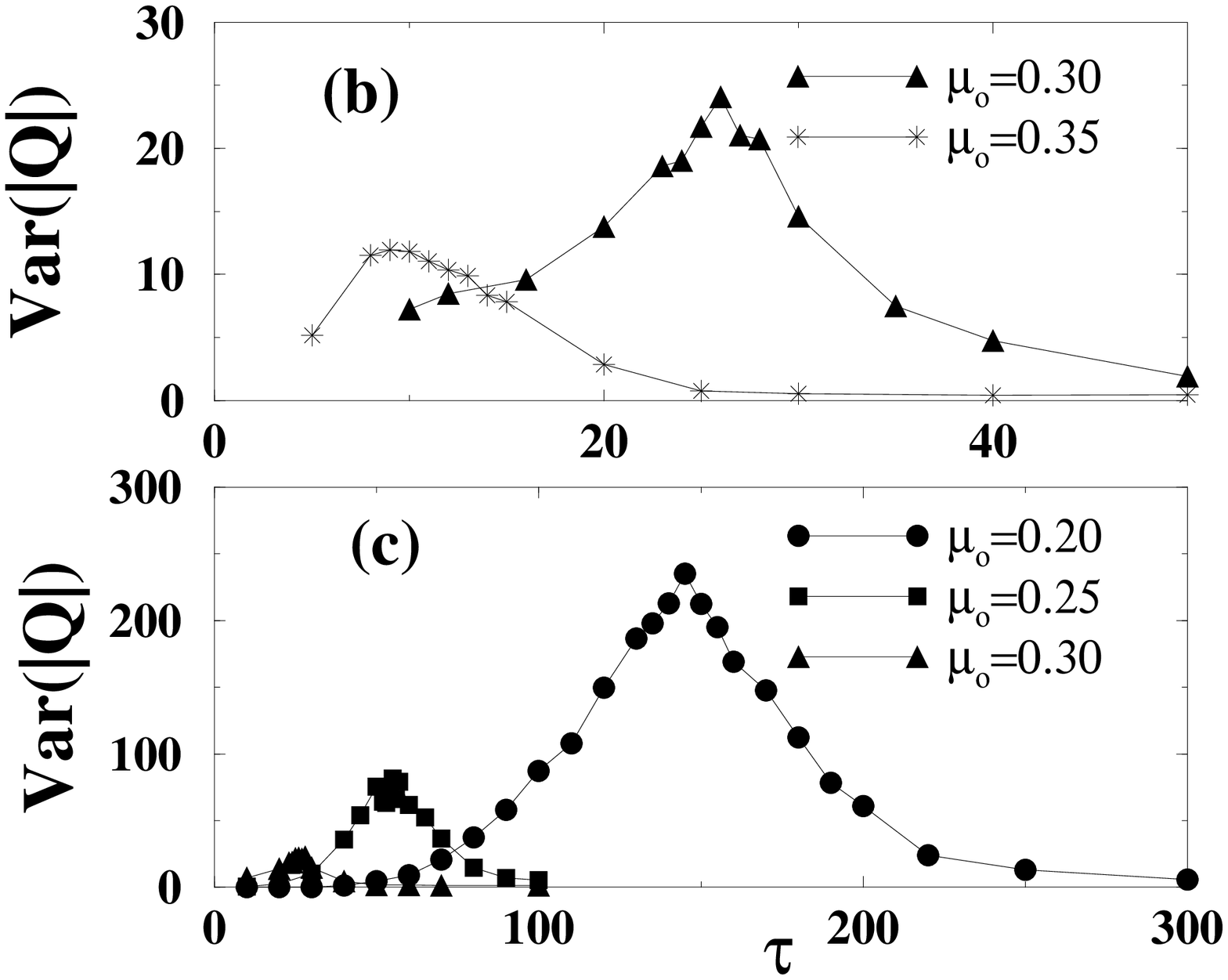}}
\caption{Plots of (a) the order parameter and (b) the fluctuations
of the order parameter versus the period $\tau$ of the applied
chemical potential, respectively.
Results obtained by using lattices of side
$L = 100$ and by taking different values of the amplitude of the
applied signal, as listed in the figures.}
\label{Fig6}
\end{figure}

\begin{figure}
\centerline{
\includegraphics[width=10cm,height=10cm,angle=0]{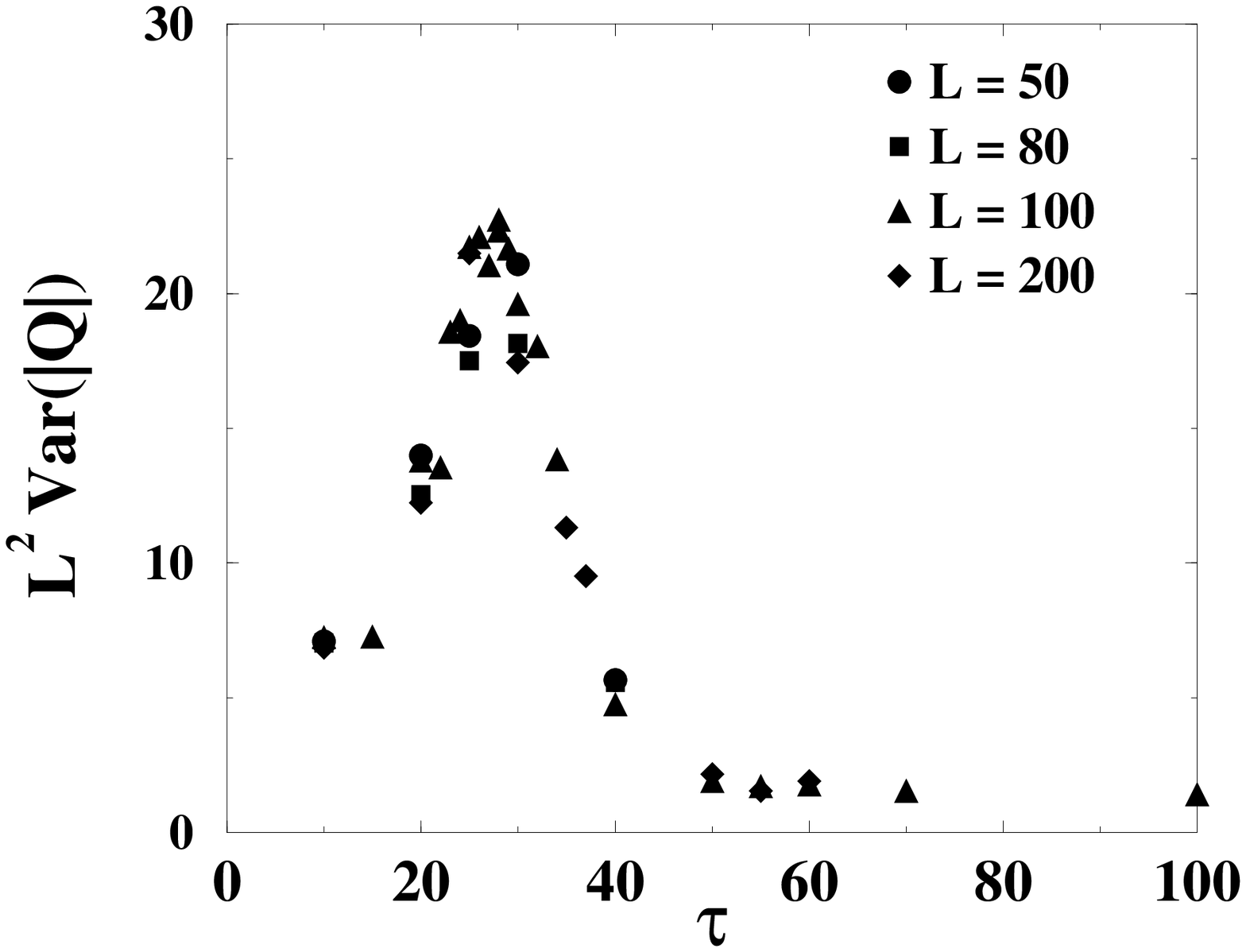}}
\caption{Plots of the fluctuations of the order parameter,
as defined by equation \ref{chiorpa}), 
versus the period $\tau$ of the applied
chemical potential.
Results obtained by using lattices of different side
as listed in the figure.}
\label{Fig7}
\end{figure}

\begin{figure}
\centerline{
\includegraphics[width=10cm,height=10cm,angle=0]{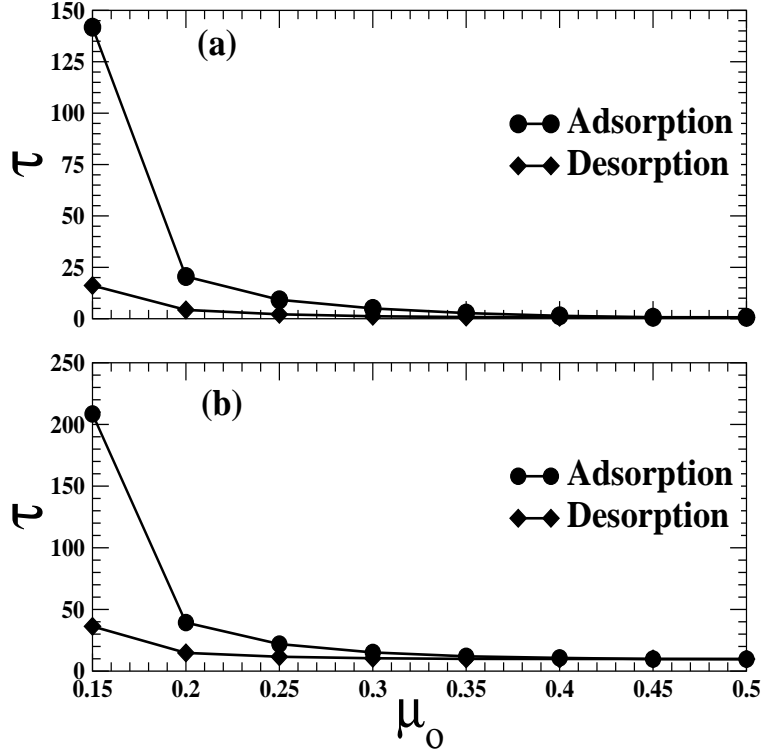}}
\caption{
Plot of the relaxation times versus the applied overpotential.
$\tau_{des}^{1/2}$ and $\tau_{des}^{0}$ are shown by means of 
full diamonds in figures a) and b), respectively. These data corresponds to 
the desorption processes up to $\theta_{Ag} = 1/2$ and $\theta_{Ag} = 0$,
respectively.
$\tau_{ads}^{1/2}$ and $\tau_{ads}^{1}$ are shown by means of 
full circles in figures a) and b), respectively.
These data corresponds to the adsorption processes up 
to $\theta_{Ag} = 1/2$ and $\theta_{Ag} = 1$, respectively.
Results obtained by averaging over $100$ different configurations.
More details in the text.
}
\label{Fig8}
\end{figure}

In view of these findings it would be desirable to perform a systematic 
study of finite-size effects. In fact, it is well known that continuous
(second-order) phase transitions become shiftted and rounded due to the
finite size of the samples used in numerical simulations, while actual 
transitions can only be observed in the thermodynamic limit    
($L \rightarrow \infty $). This shortcoming can be overcome by 
applying the finite-size scaling theory to the numerical results 
obtained by using a wide range of sample sizes \cite{103,107}. 
Therefore we performed simulations 
up to $L = 1024$ for few typical values of the
parameters (not shown here for the sake of clarity). 
Our first finding was that the dependence 
of $\langle | Q | \rangle $ on $\mu_{o}$ does not show any appreciable 
finite-size effect (not shown here for the sake of space). 
Furthermore, the fluctuations of the order 
parameter ($\chi$, as measured according to equation (\ref{chiorpa})) are 
independent of the lattice size, as shown in figure \ref{Fig7}. 
Since one has that $\chi = L^ {2} Var (Q) $, it follows that the
variance of the order parameter actually vanishes in the thermodynamic 
limit. Due to this evidence we conclude that for the range of parameters 
used, rather than undergoing a true phase 
transition, the $Ag/Au(100)$ system actually exhibits 
a crossover between the DOS and the DDS. 
This finding may be consistent with the observation that the multidomain
growth regime always prevails in our simulations (see also the 
discussion related to figures \ref{Fig3} and \ref{Fig4} ).  
In fact, one may expect 
that the random growth and subsequent dissolution of many domains, as
well as the absence of a well-defined critical nucleous, would show a 
weak (if any) size dependence of the physical observables. However, 
one has to recognize that Sides et al \cite{dro3} have found evidence
of a DPT in the kinetic Ising model only within the MD regime, altough 
the size-dependence reported in that work is rather weak.
However, in a subsequent paper Korniss et al \cite{kor} have reported 
an extensive study of the finite-size effects that can clearly be 
understood within the framework of the finite-size
scaling theory. 
On the other hand, a possible reason for the observation of a 
crossover instead of a true DPT could be the absence of symmetry
between the adsorption and desorption processes, as already pointed
out upon the qualitative discussion of the snapshots shown in figures
\ref{Fig3} and \ref{Fig4}. In order to test this possibility we 
measured the relaxation times for $Ag-$covered and uncovered surfaces
as a function of the applied overpotential (see figure \ref{Fig8}). 
Two relaxation times are measured for each process, namely:
i) $\tau_{des}^{1/2}$ and $\tau_{des}^{0}$ for the desorption
processes up to $\theta_{Ag} = 1/2$ and $\theta_{Ag} = 0$,
respectively; and ii)  $\tau_{ads}^{1/2}$ and $\tau_{ads}^{1}$ 
for the adsorption processes up to $\theta_{Ag} = 1/2$ 
and $\theta_{Ag} = 1$, respectively. The obtained results,
plotted in figure \ref{Fig8}, shown that for low overpotentials
the relaxation times corresponding to both processes are different,
quantitatively confirming the asymmetry between them.
Also, for large overpotentials the relaxation times tend to be almost 
the same, suggesting that in this limit the asymmetry may 
be irrelevant. However, if should also be noticed that in that case neither
crossovers nor DPT's are observed due to the fact that one would  need to
apply signals with periods smaller than $1 mcs$ (see figure \ref{Fig6} ).    
Within this context, it is worth mentioning that the response 
to a pulsed perturbation of 
the Ziff-Gulari-Barshad (ZGB) model \cite{zgb0} for the catalytic oxidation of 
$CO$, has also very recently been studied \cite{zgb1}.
The ZGB model exhibits a first-order irreversible phase transition
between an active state with $CO_{2}-$production and an absorbing
(or poisoned) state with the surface of the catalyst fully
covered by $CO$, such that in this regime the reaction stops 
irreversible \cite{zgb0} (for a recent review 
see e.g. reference \cite{yass}). In order to study DPT's in the ZGB model it
is convenient to perform a generalization by introducing 
a small probability for $CO$ desorption that, on the one hand
preserves the first-order nature of the transition \cite{eva},
and on the other hand prevents the occurrence of $CO-$poisoning.
Measurements of the lifetimes associated with the decay of the 
metastable states of the ZGB model indicate that they depend on the 
direction of the process, showing a marked asymmetry as in the case 
of our adsorption-desorption simulations. In fact, the contamination time
$\tau_{d}$ (measured when the system is quenched from high to low
$CO-$coverage) is different from the poisoning time  
$\tau_{p}$ (measured from low to high $CO-$coverage)\cite{zgb1,zgb2}.
Therefore, DPT's are observed by applying a periodic
external -asymmetric- signal of period 
$\tau = \tau_{p} + \tau_{d}$ \cite{zgb1}. 
Based on these evidence, we conclude that the observation of the crossover
in our simulations should -most likely- be related to the fact that we have 
applied a periodic -symmetric- potential and the involved adsorption-desorption
processes are not symmetric.
Of course, it could also be possible that at higher 
temperatures the asymmetry may become irrelevant. However, in the 
present work we have restricted ourselves to $T = 300K$ 
since we expect to contribute to the understanding of the dynamic 
behavior of the $Ag/Au(100)$ system under a standard 
electrochemical environment.  

\begin{figure}
\centerline{
\includegraphics[width=10cm,height=10cm,angle=0]{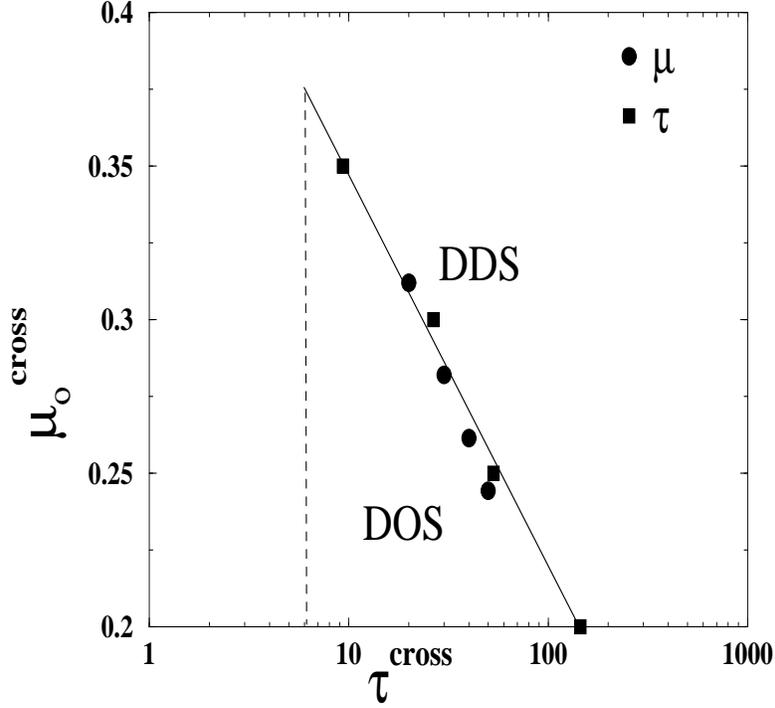}}
\caption{Linear-logarithmic plot of the crossover chemical potential 
versus the crossover period. Data obtained from the peaks observed
in plots of $Var(|Q|)$ versus $\mu_{o}$ (full circles),
and  $Var(|Q|)$ versus $\tau$ (full squares).
The full line showing the border between 
dynamic ordered states (DOS) and
dynamic disordered states (DDS) has been drawn 
in order to guide the eyes. Also, the (vertical)
dashed line corresponds to the minimum period 
at which DDS's were observed. More details in the text.}
\label{Fig9}
\end{figure}
In order to further characterize the crossover between 
different states we take advantage of the well defined 
peak exhibited by $\chi$ and $Var(|Q|)$ (see 
figures \ref{Fig5}, \ref{Fig6}, and \ref{Fig7}). 
So, the crossover period ($\tau^{cross}$) and the 
corresponding crossover chemical 
potencial ($\mu_{o}^{cross}$) are identified with 
the location of the above-mentioned peak. The obtained results 
are displayed in figure \ref{Fig9}, which shows a logarithmic 
dependence of $\mu_{o}^{cross}$ on $\tau^{cross}$. In the
``state diagram'' of figure \ref{Fig9}, the full line shows the 
border bewteen DOS's obtained for 
intermediate values of both $\mu_{o}^{cross}$ and $\tau^{cross}$,
and DDS's that are found for large 
enough values of the period. It should be noted that 
for very low periods (i.e.  $\tau^{cross} < 6$) we 
were unable to observe DOS (see e.g. figure \ref{Fig6}) 
even after largely increasing the chemical potential.
A similar effect has been 
reported by Korniss et al \cite{kor} in the case of the 
Ising model, see e.g. figure 15 of reference \cite{kor}.
Therefore, a (vertical) dashed line was drawn in the diagram 
of figure \ref{Fig9} in order to properly confine the
region where DOS's are found. An analog phase diagram 
has been reported by Korniss et al for the case of DPT's 
in the Ising model, see e.g. 
figure 16(a) of reference \cite{kor}.\\    

{\bf IV. CONCLUSIONS.}\\

The $Ag/Au(100)$ system previously simulated  by means of the Monte Carlo method
under equilibrium conditions, is analyzed in the present 
work from another point of
view, namely by studying its dynamic behavior due to the application
of an external drive.
The physical situation that motivates this paper is the adsorption
isotherm of $Ag$ on the $Au(100)$ surface, which exhibits an abrupt jump
between a low-coverage state and high-coverage state, for a well-defined
(coexistence) chemical potential at $T=300 K$. 
In this work, we analize the situation where the chemical potential
is varied periodically around $\mu_{coex}$ and
we study the influence of the period and the amplitude of that variation
on the dynamic behavior of the system.

Our results show that a silver layer adsorbed on $Au(100)$ smoothly
changes from a dynamically ordered state (DOS) to a dynamically 
disordered state (DDS) when the period (amplitude) of the chemical potential
is increased by keeping $\mu_{o}$ ($\tau$) constant. Since the size dependence
of the order parameter is negligible and its fluctuations scale 
with the system size, we conclude that the system does not exhibit a true
dynamic phase transition, but rather it exhibits a crossover between
two different dynamic states, namely DOS and DDS, respectively.
The absence of true DPT's is due to the asymmetry bewteen the 
adsorption and desorption processes of $Ag-$atoms.
The crossover points are identified by the position of the peaks of the 
fluctuations of the order parameter. In this way we are able to 
draw the corresponding diagram of states characteristic of the system.
The amplitude of the chemical potential at the crossover point exhibits a 
logarithmic dependence on the crossover period. However, for low enough periods
the DOS is no longer observed.   

The experiments that we propose in order to verify these theoretical results
consist in the study of the adsorption of silver on a gold $100$ surface, 
under Upd conditions, and the subsequent analysis of the influence of a 
periodic variation of the applied potential. One has to recognize that 
it would be difficult to establish a correlation between the actual time 
scale of the experiments and the Monte Carlo time step, although
qualitative similar observations are expected. Furthermore, by determining the 
rate constants of the relevant electrochemical processes, one may perform 
real-time Monte Carlo simulations \cite{w1}.\\ 

{\bf  ACKNOWLEDGEMENTS}. This work was financially supported by
CONICET, UNLP and  ANPCyT (Argentina).

\end{document}